\newcommand{\beq}{\begin{equation}}
\newcommand{\eeq}{\end{equation}}
\newcommand{\bea}{\begin{eqnarray}}
\newcommand{\eea}{\end{eqnarray}}

\newcommand{\gsim}{\lower.7ex\hbox{$\;\stackrel{\textstyle>}{\sim}\;$}}
\newcommand{\lsim}{\lower.7ex\hbox{$\;\stackrel{\textstyle<}{\sim}\;$}}

\newcommand{\mrm}{\mathrm}

\documentclass[a4paper]{jpconf}
\usepackage{graphicx}
\usepackage{epsf}
\usepackage{epstopdf}
\usepackage{mathrsfs}
\usepackage{amssymb}
\usepackage{verbatim}

\def\stacksymbols #1#2#3#4{\def\theguybelow{#2}
    \def\vp{\lower#3pt}
    \def\sp{\baselineskip0pt\lineskip#4pt}
    \mathrel{\mathpalette\intermediary#1}}

\def\intermediary#1#2{\vp\vbox{\sp
     \everycr={}\tabskip0pt
     \halign{$\mathsurround0pt#1\hfil##\hfil$\crcr#2\crcr
              \theguybelow\crcr}}}

\def\be{\begin{equation}}
\def\ee{\end{equation}}
\def\bea{\begin{eqnarray}}
\def\eea{\end{eqnarray}}

\def\sp{\;\;\;,\;\;\;}

\def\mrm{\mathrm}

\def\lsim{\raise0.3ex\hbox{$\;<$\kern-0.75em\raise-1.1ex\hbox{$\sim\;$}}}
\def\gsim{\raise0.3ex\hbox{$\;>$\kern-0.75em\raise-1.1ex\hbox{$\sim\;$}}}

\def\inbar{\,\vrule height1.5ex width.4pt depth0pt}

\def\IC{\relax\hbox{$\inbar\kern-.3em{\rm C}$}}
\def\IQ{\relax\hbox{$\inbar\kern-.3em{\rm Q}$}}
\def\IR{\relax{\rm I\kern-.18em R}}
 \font\cmss=cmss10 \font\cmsss=cmss10 at 7pt
\def\IZ{\relax\ifmmode\mathchoice
 {\hbox{\cmss Z\kern-.4em Z}}{\hbox{\cmss Z\kern-.4em Z}}
 {\lower.9pt\hbox{\cmsss Z\kern-.4em Z}}
 {\lower1.2pt\hbox{\cmsss Z\kern-.4em Z}}\else{\cmss Z\kern-.4em Z}\fi}

\begin{document}

\title{Invisible Higgs and Scalar Dark Matter }

\author{Yann Mambrini}

\address{Laboratoire de Physique Th\'eorique Universit\'e Paris-Sud, F-91405 Orsay, France}

\ead{yann.mambrini@th.u-psud.fr}

\begin{abstract}
In this proceeding, we show that when we combined WMAP and the most recent results of XENON100, the invisible width of the Higgs
to scalar dark matter is negligible($\lesssim 10 \%$),  except in a small region with very light dark matter ($\lesssim 10$ GeV) not yet 
excluded by XENON100 or around 60 GeV where the ratio can reach 50\% to 60\%. 
The new results released by the Higgs searches of ATLAS and CMS set very strong limits on the elastic scattering cross section.
\end{abstract}

\section{Introduction}

Two of the most important issues in particle physics phenomenology are the nature
 of the dark matter and the mechanism to
realize spontaneously the electroweak symmetry breaking of the Standard Model (SM).
 The observations made
by the WMAP collaboration show that the matter content of the universe is dark,
making up about 85 \% of the total amount of matter whereas
 the XENON collaboration recently released its constraints
on direct detection of Dark Matter . These constraints are the most 
stringent in the field nowadays, and begin to exclude a significant part of the parameter space
of the Weakly Interacting Massive Particle (WIMP) paradigm. 
On the other front, the accelerator collaborations 
ATLAS, CMS and D0/CDF \cite{ATLAS}
have obtained important results concerning the Higgs searches.
. It is obvious that the Higgs hunting
at LHC is intimately linked with measurement of elastic scattering on nucleon,
especially in Higgs-portal like models where the Higgs boson is the key particle
exchanged through annihilation/scattering processes.
It has already been showed recently that a combined LEP/TEVATRON/XENON/WMAP analysis
can restrict severely the parameter space allowed in generic constructions \cite{Mambrini:2011pw}.
In this work, we apply such analysis in the specific context of a scalar singlet dark matter extension
of the Standard Model and show that most of the region allowed by WMAP
will be excluded/probed by LHC and XENON100 by the end of next year.

\section{The Model}
The simplest extension of the SM is the addition of a real singlet scalar field.Although it is possible to generalize to scenarios with more than one singlet, the simplest
case of a single additional singlet scalar provides a useful framework to analyze
the generic implications of an augmented scalar sector to the SM.
The most general renormalizable potential involving the SM Higgs doublet $H$ and the
singlet $S$ is

\bea
{\cal L}&=& {\cal L}_{SM} + (D_{\mu}H)^\dag (D^{\mu}H) + \frac{1}{2} \mu_H^2 H^\dag H 
-\frac{1}{4}\lambda_H H^4
\nonumber
\\
&+& \frac{1}{2} \partial_{\mu}S \partial^{\mu}S
-\frac{\lambda_{S}}{4} S^4 - \frac{\mu_S^2}{2} S^2 - \frac{\lambda_{HS}}{4}S^2 H^{\dag}H
\nonumber
\\
&-&\frac{\kappa_1}{2} H^\dag H S - \frac{\kappa_3}{3}S^3 - V_0
\label{Eq:Lagrangian}
\eea 
 
 \noindent
 where $D_{\mu}$ represents the covariant derivative.

Different aspects of scalar singlet extension of the SM has already been studied in 
\cite{ScalarDM1}
whereas a nice preliminary analysis of its dark matter consequences can be found in 
\cite{Barger:2007im}. Some authors also tried to explain the DAMA and/or COGENT 
excess \cite{Andreas:2008xy}
whereas other authors probed the model by indirect searches \cite{Yaguna:2008hd},
or looked at the consequences of earlier XENON data \cite{Cai:2011kb} or at the LHC
\cite{Zerwas} or all combined \cite{Mambrini:2011ik}.

Recently the XENON100 collaboration released new data, the most stringent in the field
of Dark Matter detection\footnote{Keeping an eye on the results of COGENT collaborations,
recent works showed there exists a tension between XENON100 and COGENT \cite{Schwetz:2011xm}, 
or not \cite{Hooper:2011hd}. We thus safely 
decided not to discuss in detail the COGENT issue in our analysis}.
Moreover, recently CRESST experiment released their analysis in the low mass region \cite{Angloher:2011uu} 
and seems to converge with DAMA/LIBRA and CoGENT toward a possible light dark matter signal for a mass
aound 10 GeV \cite{Kelso:2011gd}. In the meantime,
if $m_S \lesssim M_H/2$ the invisible width decay\footnote{See the works in \cite{Hinv} for an earlier study of
invisible width of the Higgs. Moreover, during the revision of this study, several independant work confirming 
our results were published
in \cite{Raidal:2011xk} and \cite{He:2011de}} of the Higgs $H \rightarrow SS$ could 
perturbate the Higgs searches at LHC based on SM Higgs branching ratio (see Eq.\ref{Eq:HiggsWidth}). 
However, one can easily understand that there exists a tension between the direct detection measurement
and the invisible branching ratio. Indeed, for decreasing mass of DM ($m_S \lesssim 100$ GeV), 
the spin independent cross section increases. 
Quantitatively speaking, one needs to compare the invisible Higgs width ($H \rightarrow SS$)

\be
\Gamma^{{\mrm inv}}_H= \frac{\lambda_{HS}^2 M_W^2}{32 \pi g^2 M_H^2} \sqrt{M_H^2 - 4 m_S^2}
\label{Eq:HiggsWidth}
\ee

\noindent
with the spin independent scattering cross section on the proton

\be
\sigma^{SI}_{S-p} = \frac{m_p^4 \lambda_{HS}^2 (\sum_{q} f_q )^2}{16 \pi (m_p + m_S)^2 M_H^4}
\label{Eq:SigmaSI}
\ee

\noindent
Combining Eq.(\ref{Eq:HiggsWidth}) and Eq.(\ref{Eq:SigmaSI}) one obtains

\be
\frac{\Gamma_H^{\mrm Inv}}{\sigma^{SI}_{S-p}}= \frac{(m_S+m_p)^2 M_H^2 M_W^2 \sqrt{M_H^2-4 m_S^2}}{2 g^2 f^2 m_p^4}
\ee

\noindent
Some points
with high invisible width still survive. They correspond
to two distinct regions: 

\begin{itemize}
\item{ A region with very light scalar ($m_S \lesssim 10$ GeV) still not yet excluded by the precision
of XENON100 experiments due to its high threshold. This correspond to very large invisible branching 
ratio}

\item{A region with $50 \mrm{GeV} \lesssim m_S \lesssim 70$ GeV with branching ratio which can
 reach 60\% to 70 \% which is the region taken in consideration in \cite{Raidal:2011xk}}.

\end{itemize} 

\begin{figure}[h]
\begin{minipage}{20pc}
\includegraphics[width=20pc]{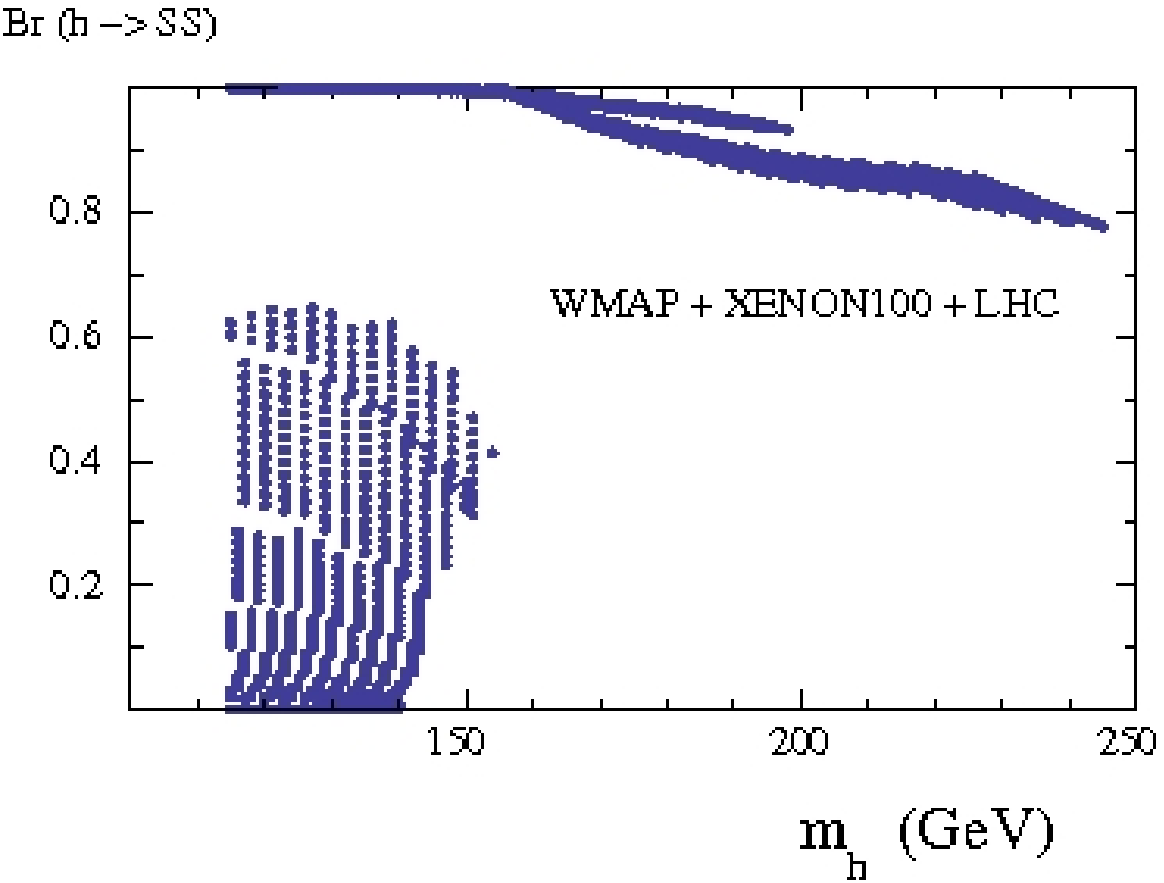}
\caption{\label{label} Branching ratio for the invisible Higgs decay into scalars  
as a function of the Higgs mass.
The scan is over ($m_h$, $m_S$, $\lambda_{hs}$) and subject to the 
WMAP, XENON100 and ATLAS/CMS constraints.    }
\label{Fig:BrInvScalHExcl1}
\end{minipage}\hspace{2pc}%
\begin{minipage}{20pc}
\includegraphics[width=20pc]{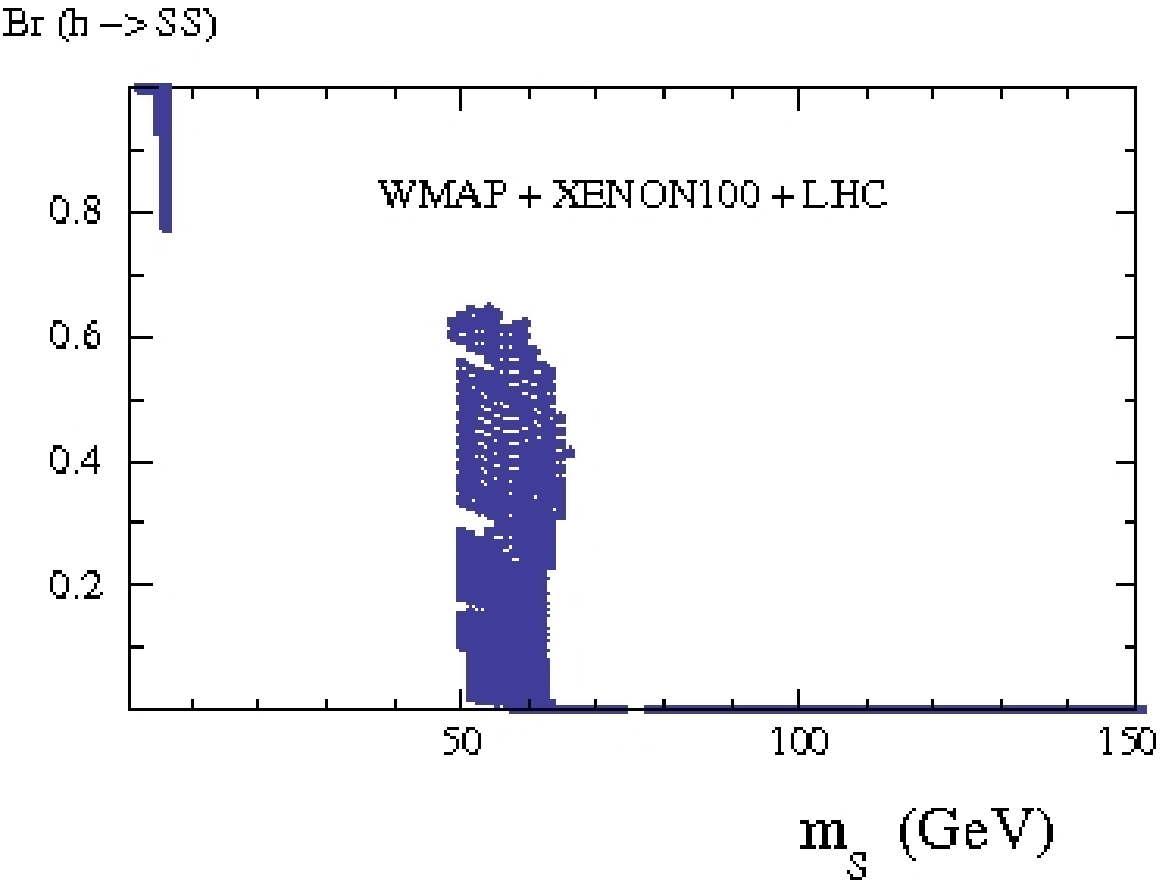}
\caption{\label{label} Branching ratio for the invisible Higgs decay into scalars  
as a function of the dark matter mass.
The scan is over ($m_h$, $m_S$, $\lambda_{hs}$) and subject to the 
WMAP, XENON100 and ATLAS/CMS constraints.  }
\label{Fig:BrInvScalHExcl2}
\end{minipage} 
\end{figure}

\noindent
We show the effects of combining WMAP and XENON100 data in Fig.\ref{Fig:BrInvScalHExcl1}
and Fig.\ref{Fig:BrInvScalHExcl2} . 
As one can see, in Fig.\ref{Fig:BrInvScalHExcl2} 
except for these two particular regions,  the majority of points respecting WMAP and XENON100 constraints give 
 very low invisible width. As a conclusion, we can 
affirm that the Higgs searches at LHC with a scalar dark matter is not affected for $m_h\gtrsim 150$ GeV
It  means that we can use the standard Higgs limit searches of ATLAS and CMS and apply them in the model
in this region of the parameter space, 
they are only slightly affected by the presence of the scalar dark matter\footnote{In our numerical study, we
obviously took into consideration the invisible Higgs width to apply the CMS and ATLAS constraints}.

Due to the last data released recently by CRESST collaboration \cite{Angloher:2011uu} it is interesting to notice that
some points in the parameter space around $m_S \simeq$ 10 GeV are not yet excluded by the latest XENON100
constraints as can be seen in the upper left corner of Fig.\ref{Fig:BrInvScalHExcl2}. These points generates
a Higgs completely invisible at the LHC.
This corresponds to the region near Br($H \rightarrow SS$) $\simeq$ 100\% in Fig.\ref{Fig:BrInvScalHExcl1}.

\noindent
It is also interesting to point out that the Higgs-portal construction is similar by several 
aspects to the Z'-portal model of dark matter \cite{Z'}:  as any Higgs searches restrict severely
the parameter space of the model, any Z' searches at LHC should be use in complementarity
with direct detection searches to probe the entire parameter space allowed by WMAP.
A vectorial Dark Matter will also give similar results \cite{VDM}
At the same time, the analysis should be done is SUSY scenario where light Higgses
are the main annihilation channel, leading to sever direct detection constraints \cite{SUSY}.
Writing the conclusion of this work, we noticed that authors just looked at some
consequences of recent Higgs searches at LHC in the NMSSM case \cite{Ellwanger:2011sk} 
and extended scalar sectors \cite{He:2011ti}.


\section*{References}
\begin{thebibliography}{9}




  
  %==============================  HIGGS LIMIT =======================================

\bibitem{ATLAS}
\begin{verbatim}
report  ATLAS-CONF-2011-112; report ATLAS-CONF-2011-157; report CMS PAS HIG-11-011; 
D0 Note 6229-CONF; report FERMILAB-CONF-10-257-E
\end{verbatim}


    \bibitem{Mambrini:2011pw}
  Y.~Mambrini, B.~Zaldivar,
  %``When LEP and Tevatron combined with WMAP and XENON100 shed light on the nature of Dark Matter,''
  [arXiv:1106.4819 [hep-ph]].


%================================   SCALAR DM MODEL   =============================================

%\cite{Silveira:1985rk}
\bibitem{ScalarDM1}
  V.~Silveira, A.~Zee,
  %``Scalar Phantoms,''
  Phys.\ Lett.\  {\bf B161}, 136 (1985);
 In the context of mirror matter, the Higgs portal interaction 
also appeared in  
 R.~Foot, H.~Lew and R.~R.~Volkas,
  %``A Model with fundamental improper space-time symmetries,''
  Phys.\ Lett.\ B\ {\bf 272}, 67  (1991);  J.~McDonald,
  %``Gauge singlet scalars as cold dark matter,''
  Phys.\ Rev.\  {\bf D50 } (1994)  3637-3649.
  [hep-ph/0702143 [HEP-PH]];  J.~McDonald,
  %``Thermally generated gauge singlet scalars as selfinteracting dark matter,''
  Phys.\ Rev.\ Lett.\  {\bf 88 } (2002)  091304.
  [hep-ph/0106249];  C.~P.~Burgess, M.~Pospelov, T.~ter Veldhuis,
  %``The Minimal model of nonbaryonic dark matter: A Singlet scalar,''
  Nucl.\ Phys.\  {\bf B619 } (2001)  709-728.
  [hep-ph/0011335];
  B.~Patt, F.~Wilczek,
  %``Higgs-field portal into hidden sectors,''
  [hep-ph/0605188];  K.~A.~Meissner, H.~Nicolai,
  %``Conformal Symmetry and the Standard Model,''
  Phys.\ Lett.\  {\bf B648 } (2007)  312-317.
  [hep-th/0612165];
  H.~Davoudiasl, R.~Kitano, T.~Li, H.~Murayama,
  %``The New minimal standard model,''
  Phys.\ Lett.\  {\bf B609 } (2005)  117-123.
  [hep-ph/0405097];  S.~-h.~Zhu,
  %``Electro-weak symmetry spontaneously breaking and cold dark matter,''
    [hep-ph/0601224];
  X.~-G.~He, T.~Li, X.~-Q.~Li, J.~Tandean, H.~-C.~Tsai,
  %``Constraints on Scalar Dark Matter from Direct Experimental Searches,''
  Phys.\ Rev.\  {\bf D79 } (2009)  023521.
  [arXiv:0811.0658 [hep-ph]];
   X.~-G.~He, T.~Li, X.~-Q.~Li, J.~Tandean, H.~-C.~Tsai,
  %``The Simplest Dark-Matter Model, CDMS II Results, and Higgs Detection at LHC,''
  Phys.\ Lett.\  {\bf B688 } (2010)  332-336.
  [arXiv:0912.4722 [hep-ph]];
  X.~-G.~He, S.~-Y.~Ho, J.~Tandean, H.~-C.~Tsai,
  %``Scalar Dark Matter and Standard Model with Four Generations,''
  Phys.\ Rev.\  {\bf D82 } (2010)  035016.
  [arXiv:1004.3464 [hep-ph]];  S.~Kanemura, S.~Matsumoto, T.~Nabeshima, N.~Okada,
  %``Can WIMP Dark Matter overcome the Nightmare Scenario?,''
  Phys.\ Rev.\  {\bf D82 } (2010)  055026.
  [arXiv:1005.5651 [hep-ph]];  M.~Aoki, S.~Kanemura, O.~Seto,
  %``Multi-Higgs portal dark matter under the CDMS II results,''
  Phys.\ Lett.\  {\bf B685 } (2010)  313-317.
  [arXiv:0912.5536 [hep-ph]];  W.~-L.~Guo, Y.~-L.~Wu,
  %``The Real singlet scalar dark matter model,''
  JHEP {\bf 1010 } (2010)  083.
  [arXiv:1006.2518 [hep-ph]];  V.~Barger, Y.~Gao, M.~McCaskey, G.~Shaughnessy,
  %``Light Higgs Boson, Light Dark Matter and Gamma Rays,''
  Phys.\ Rev.\  {\bf D82 } (2010)  095011.
  [arXiv:1008.1796 [hep-ph]];  J.~R.~Espinosa, M.~Quiros,
  %``Novel Effects in Electroweak Breaking from a Hidden Sector,''
  Phys.\ Rev.\  {\bf D76 } (2007)  076004.
  [hep-ph/0701145];
  H.~Sung Cheon, S.~K.~Kang, C.~S.~Kim,
  %``Low Scale Leptogenesis and Dark Matter Candidates in an Extended Seesaw Model,''
  JCAP {\bf 0805 } (2008)  004.
  [arXiv:0710.2416 [hep-ph]]; 
  A.~Melfo, M.~Nemevsek, F.~Nesti, G.~Senjanovic and Y.~Zhang,
  %``Inert Doublet Dark Matter and Mirror/Extra Families after Xenon100,''
  Phys.\ Rev.\ D {\bf 84} (2011) 034009
  [arXiv:1105.4611 [hep-ph]].
  %%CITATION = ARXIV:1105.4611;%%
  
  
  \bibitem{Barger:2007im}
   V.~Barger, P.~Langacker, M.~McCaskey, M.~J.~Ramsey-Musolf, G.~Shaughnessy,
  %``LHC Phenomenology of an Extended Standard Model with a Real Scalar Singlet,''
  Phys.\ Rev.\  {\bf D77 } (2008)  035005.
  [arXiv:0706.4311 [hep-ph]];
    T.~E.~Clark, B.~Liu, S.~T.~Love, T.~ter Veldhuis,
  %``The Standard Model Higgs Boson-Inflaton and Dark Matter,''
  Phys.\ Rev.\  {\bf D80}, 075019 (2009),
  [arXiv:0906.5595 [hep-ph]];
R.~N.~Lerner, J.~McDonald,
  %``Gauge singlet scalar as inflaton and thermal relic dark matter,''
  Phys.\ Rev.\  {\bf D80}, 123507 (2009),
  [arXiv:0909.0520 [hep-ph]];
O.~Lebedev, H.~M.~Lee,
  %``Higgs Portal Inflation,''
    [arXiv:1105.2284 [hep-ph]].






\bibitem{Andreas:2008xy}
  S.~Andreas, T.~Hambye, M.~H.~G.~Tytgat,
  %``WIMP dark matter, Higgs exchange and DAMA,''
  JCAP {\bf 0810 } (2008)  034.
  [arXiv:0808.0255 [hep-ph]];
    S.~Andreas, C.~Arina, T.~Hambye, F.~-S.~Ling, M.~H.~G.~Tytgat,
  %``A light scalar WIMP through the Higgs portal and CoGeNT,''
  Phys.\ Rev.\  {\bf D82 } (2010)  043522.
  [arXiv:1003.2595 [hep-ph]];
    M.~H.~G.~Tytgat,
  %``A light scalar WIMP through the Higgs portal?,''
  [arXiv:1012.0576 [hep-ph]].



\bibitem{Yaguna:2008hd}
  C.~E.~Yaguna,
  %``Gamma rays from the annihilation of singlet scalar dark matter,''
  JCAP {\bf 0903 } (2009)  003.
  [arXiv:0810.4267 [hep-ph]];
    A.~Goudelis, Y.~Mambrini, C.~Yaguna,
  %``Antimatter signals of singlet scalar dark matter,''
  JCAP {\bf 0912 } (2009)  008.
  [arXiv:0909.2799 [hep-ph]];
    C.~Arina, M.~H.~G.~Tytgat,
  %``Constraints on Light WIMP candidates from the Isotropic Diffuse Gamma-Ray Emission,''
  JCAP {\bf 1101 } (2011)  011.
  [arXiv:1007.2765 [astro-ph.CO]].
  
  \bibitem{Cai:2011kb}
  Y.~Cai, X.~-G.~He, B.~Ren,
  %``Low Mass Dark Matter and Invisible Higgs Width In Darkon Models,''
  Phys.\ Rev.\  {\bf D83 } (2011)  083524.
  [arXiv:1102.1522 [hep-ph]]; 
   A.~Biswas, D.~Majumdar,
  %``The Real Gauge Singlet Scalar Extension of Standard Model: A Possible Candidate of Cold Dark Matter,''
    [arXiv:1102.3024 [hep-ph]];
      M.~Farina, M.~Kadastik, D.~Pappadopulo, J.~Pata, M.~Raidal, A.~Strumia,
  %``Implications of XENON100 results for Dark Matter models and for the LHC,''
   [arXiv:1104.3572 [hep-ph]].


%\cite{arXiv:1106.3097}
\bibitem{Zerwas} 
  C.~Englert, T.~Plehn, D.~Zerwas and P.~M.~Zerwas,
  %``Exploring the Higgs portal,''
  Phys.\ Lett.\ B\ {\bf 703}, 298  (2011)
  [arXiv:1106.3097 [hep-ph]];
  %%CITATION = PHLTA,B703,298;%%
   K.~Ghosh, B.~Mukhopadhyaya and U.~Sarkar,
  %``Signals of an invisibly decaying Higgs in a scalar dark matter scenario: a study for the Large Hadron Collider,''
  Phys.\ Rev.\ D {\bf 84} (2011) 015017
  [arXiv:1105.5837 [hep-ph]].
  %%CITATION = ARXIV:1105.5837;%%
   C.~Englert, T.~Plehn, M.~Rauch, D.~Zerwas and P.~M.~Zerwas,
  %``LHC: Standard Higgs and Hidden Higgs,''
  arXiv:1112.3007 [Unknown].
  %%CITATION = ARXIV:1112.3007;%%


 \bibitem{Mambrini:2011ik}
  Y.~Mambrini,
  %``Higgs searches and singlet scalar dark matter: Combined constraints from XENON 100 and the LHC,''
  arXiv:1108.0671 [hep-ph];
  %%CITATION = ARXIV:1108.0671;%%
 A.~Djouadi, O.~Lebedev, Y.~Mambrini and J.~Quevillon,
  %``Implications of LHC searches for Higgs--portal dark matter,''
  arXiv:1112.3299 [hep-ph];
  %%CITATION = ARXIV:1112.3299;%%
  M.~Kadastik, K.~Kannike, A.~Racioppi and M.~Raidal,
  %``Implications of 125 GeV Higgs boson on scalar dark matter and on the CMSSM phenomenology,''
  arXiv:1112.3647 [hep-ph];
  %%CITATION = ARXIV:1112.3647;%%



%===========================================================================================










\bibitem{Schwetz:2011xm}
  T.~Schwetz, J.~Zupan,
  %``Dark Matter attempts for CoGeNT and DAMA,''
  [arXiv:1106.6241 [hep-ph]];
  M.~T.~Frandsen, F.~Kahlhoefer, J.~March-Russell, C.~McCabe, M.~McCullough, K.~Schmidt-Hoberg,
  %``On the DAMA and CoGeNT Modulations,''
  
  \bibitem{Hooper:2011hd}
  D.~Hooper, C.~Kelso,
  %``Implications of CoGeNT's New Results For Dark Matter,''
   [arXiv:1106.1066 [hep-ph]].
   
   
   \bibitem{Angloher:2011uu}
  G.~Angloher, M.~Bauer, I.~Bavykina, A.~Bento, C.~Bucci, C.~Ciemniak, G.~Deuter, F.~von Feilitzsch {\it et al.},
  %``Results from 730 kg days of the CRESST-II Dark Matter Search,''
   [arXiv:1109.0702 [astro-ph.CO]].

\bibitem{Kelso:2011gd}
  C.~Kelso, D.~Hooper, M.~R.~Buckley,
  %``Toward A Consistent Picture For CRESST, CoGeNT and DAMA,''
   [arXiv:1110.5338 [astro-ph.CO]];
    J.~Kopp, T.~Schwetz, J.~Zupan,
  %``Light Dark Matter in the light of CRESST-II,''
  [arXiv:1110.2721 [hep-ph]].


%================================   INVISIBLE HIGGS  ============================


\bibitem{Hinv}
  M.~C.~Bento, O.~Bertolami, R.~Rosenfeld,
  %``Cosmological constraints on an invisibly decaying Higgs boson,''
  Phys.\ Lett.\  {\bf B518 } (2001)  276-281
  [hep-ph/0103340];
   M.~C.~Bento, O.~Bertolami, R.~Rosenfeld, L.~Teodoro,
  %``Selfinteracting dark matter and invisibly decaying Higgs,''
  Phys.\ Rev.\  {\bf D62 } (2000)  041302.
  [astro-ph/0003350];
  R. E. Shrock and M. Suzuki, "Invisible Decays of Higgs Bosons", Phys. Lett. 
110B, 250 (1982);
  D.~A.~Demir,
  %``Gravity wraps Higgs boson,''
  arXiv:1110.3815 [hep-ph];
  %%CITATION = ARXIV:1110.3815;%%
  
 
  
  \bibitem{Raidal:2011xk}
  M.~Raidal, A.~Strumia,
  %``Hints for a non-standard Higgs boson from the LHC,''
   [arXiv:1108.4903 [hep-ph]].
  
  \bibitem{He:2011de}
  X.~-G.~He, J.~Tandean,
  %``Light Dark Matter and Hidden Higgs Boson at the LHC,''
    [arXiv:1109.1277 [hep-ph]].
  












%-------------- vector DM -------------------------- %


%\cite{Hambye:2008bq}
\bibitem{VDM}
  T.~Hambye,
  %``Hidden vector dark matter,''
  JHEP {\bf 0901}, 028 (2009)
  [arXiv:0811.0172 [hep-ph]];
  %%CITATION = JHEPA,0901,028;%%
T.~Hambye and M.~H.~G.~Tytgat,
  %``Confined hidden vector dark matter,''
  Phys.\ Lett.\  B {\bf 683}, 39 (2010)
  [arXiv:0907.1007 [hep-ph]];
    J.~Hisano, K.~Ishiwata, N.~Nagata and M.~Yamanaka,
  %``Direct Detection of Vector Dark Matter,''
  Prog.\ Theor.\ Phys.\  {\bf 126}, 435 (2011)
  [arXiv:1012.5455 [hep-ph]];
    O.~Lebedev, H.~M.~Lee and Y.~Mambrini,
  %``Vector Higgs-portal dark matter and the invisible Higgs,''
  arXiv:1111.4482 [hep-ph].
  %%CITATION = ARXIV:1111.4482;%%









%==============================  Z'  =======================================


\bibitem{Z'}
 Y.~Mambrini,
  %``The ZZ' kinetic mixing in the light of the recent direct and indirect dark matter searches,''
  JCAP {\bf 1107 } (2011)  009.
  [arXiv:1104.4799 [hep-ph]];
   M.~T.~Frandsen, F.~Kahlhoefer, S.~Sarkar, K.~Schmidt-Hoberg,
  %``Direct detection of dark matter in models with a light Z',''
  [arXiv:1107.2118 [hep-ph]];
    D.~Feldman, B.~Kors, P.~Nath,
  %``Extra-weakly Interacting Dark Matter,''
  Phys.\ Rev.\  {\bf D75 } (2007)  023503.
  [hep-ph/0610133];
  Y.~Mambrini,
  ``The Kinetic dark-mixing in the light of CoGENT and XENON100,''
  JCAP {\bf 1009}, 022 (2010)
  [arXiv:1006.3318 [hep-ph]];
  %%CITATION = JCAPA,1009,022;%%
      Y.~Mambrini,
  %``A Clear Dark Matter gamma ray line generated by the Green-Schwarz mechanism,''
  JCAP {\bf 0912 } (2009)  005.
  [arXiv:0907.2918 [hep-ph]];
   K.~Cheung, J.~Song,
  %``Baryonic Z' Explanation for the CDF Wjj Excess,''
  Phys.\ Rev.\ Lett.\  {\bf 106 } (2011)  211803.
  [arXiv:1104.1375 [hep-ph]];
   E.~Dudas, Y.~Mambrini, S.~Pokorski, A.~Romagnoni,
  %``(In)visible Z-prime and dark matter,''
  JHEP {\bf 0908 } (2009)  014.
  [arXiv:0904.1745 [hep-ph]].


%=============================   SUSY   ======================================

\bibitem{SUSY}
  D.~Das, A.~Goudelis, Y.~Mambrini,
  %``Exploring SUSY light Higgs boson scenarios via dark matter experiments,''
  JCAP {\bf 1012 } (2010)  018.
  [arXiv:1007.4812 [hep-ph]];
   S.~Bhattacharya, U.~Chattopadhyay, D.~Choudhury, D.~Das, B.~Mukhopadhyaya,
  %``Non-universal scalar mass scenario with Higgs funnel region of SUSY dark matter: A Signal-based analysis for the Large Hadron Collider,''
  Phys.\ Rev.\  {\bf D81 } (2010)  075009.
  [arXiv:0907.3428 [hep-ph]];
    A.~Djouadi, Y.~Mambrini,
  %``The Higgs intense-coupling regime in constrained SUSY models and its astrophysical implications,''
  JHEP {\bf 0612 } (2006)  001.
  [hep-ph/0609234].


%===========================================================================================

\bibitem{Ellwanger:2011sk}
  U.~Ellwanger,
  %``Higgs Bosons in the Next-to-Minimal Supersymmetric Standard Model at the
  %LHC,''
  arXiv:1108.0157 [hep-ph].
  %%CITATION = ARXIV:1108.0157;%%
  
  
  \bibitem{He:2011ti}
  X.~G.~He and G.~Valencia,
  %``An extended scalar sector to address the tension between a fourth
  %generation and Higgs searches at the LHC,''
  arXiv:1108.0222 [hep-ph].
  %%CITATION = ARXIV:1108.0222;%%


  



\end{thebibliography}
\end{document}